\documentclass[12pt]{article}
\usepackage{color,multicol}
\usepackage{amsmath}
\usepackage{amssymb}
\usepackage{epsfig}
\usepackage{graphicx}
\usepackage{epstopdf}

\newcommand{\set}[1]{{\mathbb{#1}}}

\begin{document}

\title{Model Risk Analysis via Investment Structuring}
\author{Andrei N. Soklakov\footnote{Head of Equities Model Risk and Analytics, Deutsche Bank.\newline
{\sl The views expressed herein should not be considered as investment advice or promotion. They represent personal research of the author and do not necessarily reflect the view of his employers, or their associates or affiliates.} Andrei.Soklakov@(db.com, gmail.com).}}
\date{25 July 2015}
\maketitle

\begin{center}
\parbox{14cm}{
{\small
\lq\lq What are the origins of risks?'' and \lq\lq How material are they?'' -- these are the two most fundamental questions of any risk analysis. Quantitative Structuring -- a technology for building financial products -- provides economically meaningful answers for both of these questions. It does so by considering risk as an investment opportunity. The structure of the investment reveals the precise sources of risk and its expected performance measures materiality. We demonstrate these capabilities of Quantitative Structuring using a concrete practical example -- model risk in options on vol-targeted indices.
} }
\end{center}

\section{Introduction}
Success of individual businesses as well as entire industries is ultimately determined by the success of their products. Motivated by this observation and by the clear present need to improve the financial industry, we embarked on a project of Quantitative Structuring~\cite{Soklakov_2011,Soklakov_2014_WQS}.

Although Quantitative Structuring is firmly focused on product design, its techniques turn out to be useful beyond their original purpose. Here, for instance, we describe applications to model risk.

What could product design possibly tell us about model risk? Try to market a skew or a volatility product (investment or hedging) without mentioning either skew or volatility and the deep connection between product design and modeling becomes obvious. Modeling concepts deeply permeate product design. It should therefore come as no surprise that progress in structuring reflects back on modeling.

Model risk has many distinct flavors: model choice, calibration quality, implementation -- to name just a few. Although our methodologies are very general, we decided to introduce them by working through a concrete practical example. The relevant general concepts are introduced along the way as needed. The specific example we selected for this presentation is model choice for options on vol-targeted indices. Following the example, we show how the same methods are used to handle other flavors of model risk.

\newpage

\section{Example: Options on vol-targeted indices}
\subsection{Rationale}
Volatility is both a measure and a source of financial risks. Naturally, many investors are attracted to ideas which aim for stable, predicable exposure to volatility. One such idea is volatility targeting, also known as volatility control.

Let $S_t$ be the value of some index at time $t$. A vol-targeted version of this index, $X_t$, is defined by the initial condition $X_0=S_0$ and an iterative relationship such as
\begin{equation}\label{Eq:VolTarget}
\frac{X_{i+1}-X_i}{X_i}=\frac{\sigma}{\sigma_i}\cdot\frac{S_{i+1}-S_i}{S_i}\,,
\end{equation}
where $\sigma_i$ is the volatility of the original index $S_i$ at time $i$ and $\sigma$ is the volatility target, i.e. the desired volatility of the new index $X_t$.

Legal practicalities demand an indisputably clear procedure for computing $\sigma_i$. The usual estimator of realized variance over the window of $n$ most recent business days is a simple natural choice:
\begin{equation}\label{Eq:RV}
    \sigma_i=\sqrt{\frac{A}{n}\sum_{k=i-n}^i\ln^2\frac{S_k}{S_{k-1}}}\,,
\end{equation}
where $A$ is the annualization factor (the total number of business days per year).

As a compromise between the accuracy and the time-lag of the estimator, $n$ is usually chosen to be around 40 business days. With such a small data set, it is very tempting to experiment with efficiency. Different ad-hoc schemes combining different estimators have been proposed. This has lead to a number of practical variations of the vol-targeting scheme. There are of course no clear winners among such variations, so for the purpose of our demonstrations we chose to focuse on the core idea of vol-targeting as captured by Eqs.~(\ref{Eq:VolTarget}) and~(\ref{Eq:RV}) with $n=40$.

Imagine now a structuring department promoting a new investment strategy which they packaged for marketing purposes as an index. The volatility information on such an index is patchy due to its bespoke nature, so volatility targeting is used. Clients like the idea but ask for some additional capital protection. In other words, the clients want options on $X_t$.

\subsection{Model choice}

What model shall we use for pricing vanillas on $X_t$? Practitioners familiar with vol-targeting would easily recognize the proposal to use a constant volatility (Black-Scholes) model with the spot $X_0=S_0$ and the volatility equal to the target value $\sigma$. Let us imagine that we just encountered this proposal for the first time. We need to respond to the proposal but also form an independent data-driven opinion on the minimal modeling requirements for such trades.

For the sake of clarity let us choose $S_t$ to be a real index. Let us take Euro Stoxx 50 as of 28 January 2011 -- the same market data as we used for other completely unrelated illustrations of Quantitative Structuring. Furthermore, let us set the target $\sigma=10\%$. This is a typical target volatility level which is, as usual, significantly lower than the native volatility of the underlying index.

We consider a broad market of alternative bookings. To this end we implemented the volatility targeting scheme~(\ref{Eq:VolTarget})-(\ref{Eq:RV}) in Monte-Carlo (100\,000 paths) using the following diffusions for $S_t$.
\begin{itemize}
    \item ${\rm BS}_{10\%}^{\rm(target)}$ -- this is just the simplest benchmark model assuming $S_t$ follows lognormal process with constant volatility $\sigma=10\%$. Should $S_t$ indeed follow this process, volatility-targeting would be redundant.
    \item  ${\rm BS}_{23.95\%}^{\rm(ATMF)}$ -- same as above except the volatility is set to the 6 months ATM Forward value which happened to be 23.95\%.
    \item Local Vol -- the standard local volatility model~\cite{Dupire_1994,DermanKani_1994} calibrated to the actual market of Euro Stoxx vanillas.
    \item ${\rm SLV}^{\,\,\rho}_{\eta/\kappa}$ -- this is a stochastic local volatility model. We defer all relevant descriptions of this model to Appendix~\ref{sec:SLV}. Here we mention that this is a popular generalization of the local volatility model which introduces an extra stochastic factor for the volatility. Parameters $\eta$ and $\kappa$, known as the vol-of-vol and mean-reversion, determine the evolution of this extra factor and $\rho$ stands for its correlation with the equity.

        Why would we consider ${\rm SLV}^{\,\,\rho}_{\eta/\kappa}$? Looking at~Eq.~(\ref{Eq:RV}) we note that options on $X_t$ might be viewed as derivatives on volatility of $S_t$. Furthermore, looking at~Eq.~(\ref{Eq:VolTarget}) we see that the vol-spot correlation may also play a role. ${\rm SLV}^{\,\,\rho}_{\eta/\kappa}$ is a standard model which captures such effects (vol-of-vol and forward skew).

        We use several instances of this model, starting with ${\rm SLV}^{-70\%}_{5/10}$ and ${\rm SLV}^{-70\%}_{2.25/10}$. The first of these instances is a very stressed example. It can happen when the model is forced to match a really stressed and illiquid cliquet market (when the prices are heavily influenced by bid-offer spreads and cannot really be explained by the diffusions of Appendix~\ref{sec:SLV}). Most practitioners would say that such parameter values are far too extreme and the model is breaking. ${\rm SLV}^{-70\%}_{2.25/10}$ provides a bit more realistic scenario. We also consider parameters from~\cite{Madan_2007} to gain further insight into the stability of our conclusions regarding model choice.
\end{itemize}

\subsection{Model risk analysis}

Good booking proposals faithfully reflect the market view on the trade and introduce no further views of their own. When that is not the case, different booking proposals can be traded either against each other or against the market to create investment opportunities. Above we introduced a flat-vol proposal for booking options on vol-targeted indices and listed a range of possible market alternatives. Imagine now an investor who believes in this booking proposal and would like to use it as a sole basis for his investment strategy. What would this investment strategy look like and how would it perform?

Quantitative Structuring provides us with the answers. Let us arrive at them step-by-step. The first thing we have to do is to identify the underlying variable for the investment. For instance, if our investor believes that the booking proposal is especially great for vanillas on the value of the vol-targeted index in 6 months, then the variable in question is $X_{\rm 6m}$.

Now that we have a variable, we can extract the knowledge about this variable as seen by the investor and by the market. Mathematically this means extracting investor-believed and market-implied probability distributions. This is done as follows.

For a Monte-Carlo booking the investor-believed probability distribution for $X_{\rm 6m}$ is just a histogram of the simulated values $X_{\rm 6m}$. Let us imagine that the histogram has $N$ buckets and let $b_1,\dots,b_N$ be the realized frequencies which define the investor-believed distribution over these buckets.

The market-implied distribution for our chosen variable $X_{\rm 6m}$ is, again, just a histogram $m_1,\dots,m_N$ obtained by grouping the values of $X_{\rm 6m}$ over the same set of buckets, only this time we compute the values by implementing faithfully the vol-targeting mechanism~(\ref{Eq:VolTarget})-(\ref{Eq:RV}) and by using market dynamics.
\begin{center}
    \rule[1ex]{.5\textwidth}{.5pt}
\end{center}
{\sl
Given investor-believed and market-implied distributions $\{b_i\}_{i=1}^N$ and $\{m_i\}_{i=1}^N$ we can show (see Appendix~\ref{sec:Performance}) that the maximum expected rate of return is achieved by a payoff function which is proportional to
\begin{equation}\label{Eq:f=b/m}
f_i=b_i/m_i\,,
\end{equation}
and the expected rate of return from this investment is
\begin{equation}\label{Eq:ER}
{\rm ER}=\sum_{i=1}^Nb_i\ln f_i+{\rm RFR}-{\rm CR}\,,
\end{equation}
where ${\rm RFR}$ is the risk-free rate and ${\rm CR}$ is the rate charged by the market makers as commission (we encounter ${\rm CR}$ as trading costs).

The above equations tell us the maximum expected performance~(\ref{Eq:ER}) and the corresponding investment structure~(\ref{Eq:f=b/m}). Below we show how this is used to understand the materiality and the specific origins of model risk.

The risky part of ${\rm ER}$, lets call it ${\rm MRP}$ for model risk premium\footnote{In general, this quantity should really be called Investment Risk Premium. In earlier presentations on model risk we used to call it Model Risk Return (e.g.\ the original version arXiv:\/1304.7533v1 of \cite{Soklakov_2013a}).},
is given by the relative entropy
\begin{equation}\label{Eq:MRR}
{\rm MRP}=\sum_{i=1}^Nb_i\ln f_i = \sum_{i=1}^Nb_i\ln \frac{b_i}{m_i}\,.
\end{equation}
This fact, which has the same mathematical roots as Kelly's game-theoretic interpretation of entropy~\cite{Kelly_1956}, is interesting to us because of useful fundamental properties of entropy. In particular, we have Gibbs' inequality: ${\rm MRP}\geq 0$ with ${\rm MRP=0}$ if and only if the distributions $\{b_i\}$ and $\{m_i\}$ are identical. Gibbs' inequality guarantees that ${\rm MRP}$ is capturing every difference between the distributions -- from Monte-Carlo noise to conceptual disagreements such as presence or absence of skew. Nothing is going to escape our attention.

The economic implications of Eq.~(\ref{Eq:ER}) are just as fundamental. We note that the above investment would not make any economic sense if ${\rm ER} < {\rm RFR}$. Indeed, the growth-optimizing investment strategy~(\ref{Eq:f=b/m}) is very risky. No sane person would choose to go for such a risky investment if its expected rate of return was below risk-free. We derive that for the model risk to be economically material we must have at least
\begin{equation}\label{Eq:Materiality}
    {\rm MRP}>{\rm CR}\,.
\end{equation}
Note that the above materiality criterion does not depend on the details behind the concept of risk-free rate -- it drops out as long as we use it in a consistent way.

With a moment of thought we also discover that the requirement~(\ref{Eq:Materiality}) is very lenient. ${\rm MRP}$ is the only positive contribution which might push ${\rm ER}$ above ${\rm RFR}$. Consequently, for the investment to be successful, ${\rm MRP}$ has to cover more than the trading costs (CR), it has to cover all costs of running the business -- salaries, rent, etc.

The overall costs of running a business are specific to each business. At the same time these costs are very intuitive -- most business people can judge if $r$\% per annum amounts to a good investment for them or is simply not worth their efforts (see Appendix~\ref{sec:Performance}). For the purposes of our illustrations, we set a Spartan threshold of 1\% per annum. No equity derivative business would plan to survive on a smaller risky return on their operating capital. ${\rm MRP}$ below that threshold would indicate economically immaterial model risk.}

\begin{center}
    \rule[1ex]{.5\textwidth}{.5pt}
\end{center}

Let us now come back to our example of vanilla options on a vol-targeted index. So far we translated the proposed booking into the investor-believed distribution and used a set of alternative models to give us the range of market-implied distributions. Our next step is to compute ${\rm MRP}$ and judge the materiality of the corresponding model risk. Bold numbers in Fig.~1A are the annualized values of ${\rm MRP}$. We see that the model risk of using the proposed naive booking is immaterial (0.44\% p.a.) only in the case when the underlying $S_t$ follows ${\rm BS}_{10\%}^{\rm(target)}$, i.e.\ when the vol-targeting is completely redundant. Even when $S_t$ follows a Black-Scholes dynamics with some reasonable volatility value which does not coincide with the target level we have material model risk (4.92\% p.a.).
\begin{figure}
\includegraphics[width=\textwidth]{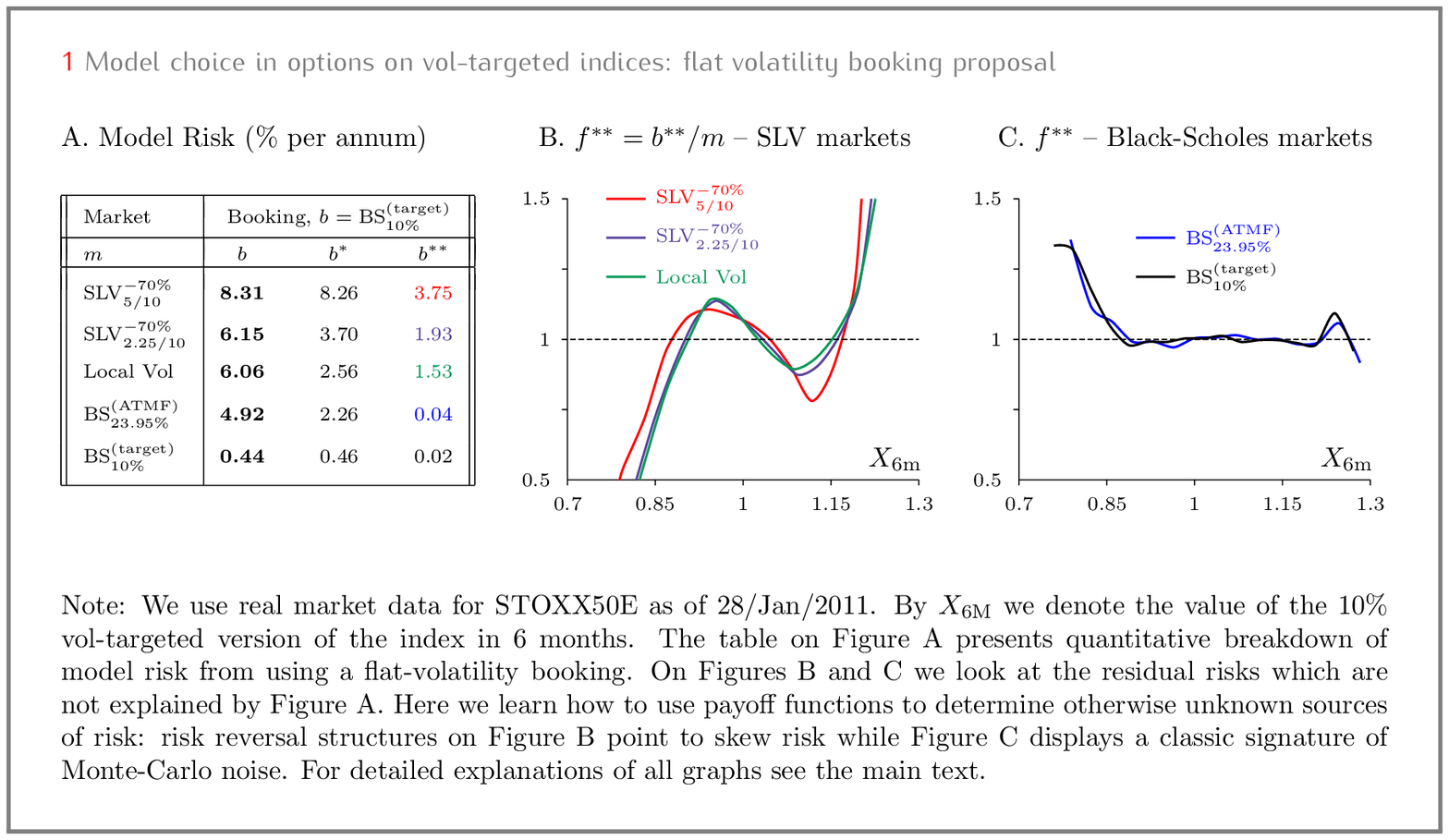}
\end{figure}
The value of ${\rm MRP}$ goes up to 6.06\% p.a.\ for the local volatility and even further to 8.31\% p.a.\ for ${\rm SLV}^{-70\%}_{5/10}$. This means that the original proposal of using flat volatility model is severely deficient. Even if we discard the results for the extremely stressed markets (represented by ${\rm SLV}^{-70\%}_{5/10}$), it is clear that, at the very least, we should use the local volatility model.

Without getting entangled in formalities, let us introduce a very practical concept of a \lq\lq good modeling choice''. Intuitively, this is given by the simplest model with the property that reasonable generalizations of it do not reveal material increase in model risk.

Looking through the table of Fig.~1A, we see that, in terms of the total ${\rm MRP}$, the difference between Local Vol and ${\rm SLV}^{-70\%}_{2.25/10}$ is not material (6.06\% vs 6.15\%). This could be an indication that the local volatility is a good modeling choice. To confirm, we need to double-check that these modeling approaches are not just equidistant from the naive flat-vol booking but are indeed close to each other. Upgrading the booking proposal to use local volatility and keeping ${\rm SLV}^{-70\%}_{2.25/10}$ as the market we compute ${\rm MRP}$ 0.25\% p.a. The difference between these two models is clearly immaterial.\footnote{Please note that we do not discuss introduction of caps (explicit or implicit) on the leverage ratio $\sigma/\sigma_i$. Depending on the contract, such caps can lead to material importance of stochastic volatility models.\label{ftn:caps}}

The ability of stochastic local volatility models to capture real markets is a topic in its own right which lies outside the scope of this demonstration. Having said that, our discussion of a good modeling choice does need some reasonable set of SLV parameters (in addition to the single cliquet-inspired instance ${\rm SLV}^{-70\%}_{2.25/10}$ considered above). We decided to use Ref.~\cite{Madan_2007} as an independent source of SLV parameters. In~\cite{Madan_2007} the authors demonstrated the capabilities of SLV model in relation to volatility derivatives. This is a good alternative to what we considered so far. As a further enhancement, we also examine non-zero correlation ($\rho=-70\%$) alongside the zero correlation used in~\cite{Madan_2007}. Results are presented on Fig.~2. We see that a lot of reasonable deviations from local volatility booking do not lead to material model risk.\footnotemark[2]
\begin{figure}
\includegraphics[width=\textwidth]{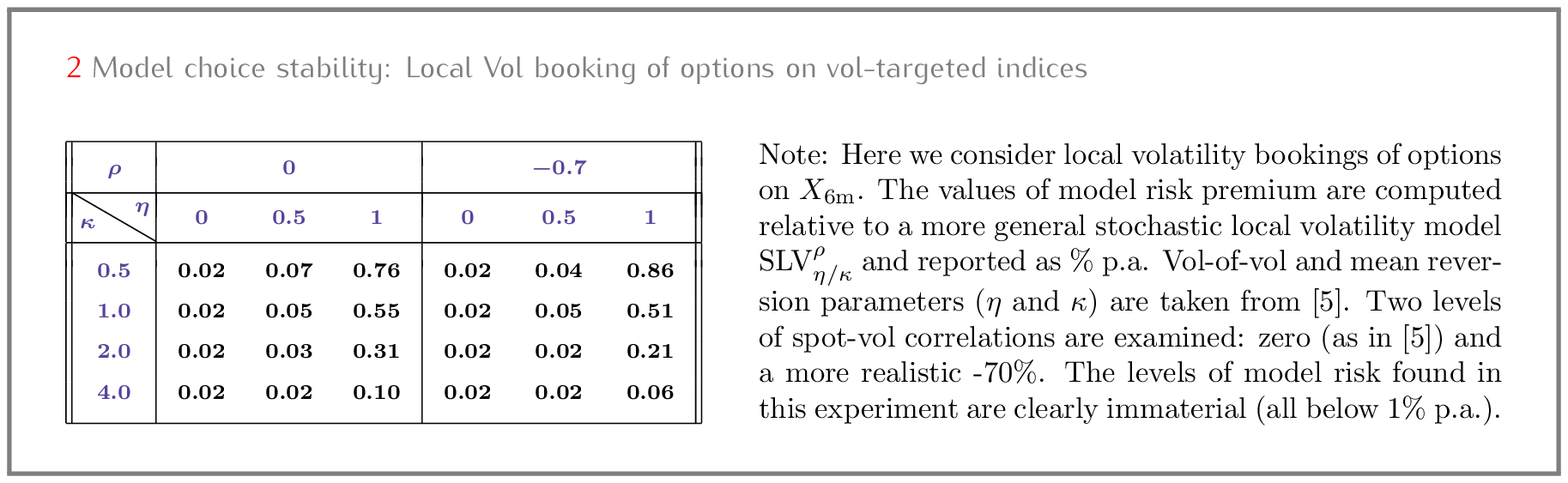}
\end{figure}
In the context of the above experiments, Local Volatility model appears to be a good modeling choice. Stochastic local volatility may be required only if you believe in the really extreme market dynamics (as captured by ${\rm SLV}^{-70\%}_{5/10}$). Although it would be prudent to keep reserves against extreme markets, the usage of models with such unrealistic dynamics for hedging purposes should be avoided.

As we already mentioned, any risk analysis revolves around two fundamental questions -- regarding the origins and the materiality of risk. So far in our example we investigated the question of overall materiality. We found that the naive booking proposal comes with model risk of at least 6\% per annum -- a material figure. We now want to determine the origins of this risk. In other words, let us investigate what exactly is wrong with the proposed booking.
\begin{center}
    \rule[1ex]{.5\textwidth}{.5pt}
\end{center}
{\sl
Let $\set{M}$ and $\set{B}$ be two sets of numbers. To make it more specific, let us think of these as the actual Monte-Carlo samples which we used to build the distributions $\{m_i\}$ and~$\{b_i\}$ in the above experiments. We found that the quality of the numbers $\set{B}$ is not good and we would like to make them statistically closer to $\set{M}$. The obvious first thing to do is to correct the mean, i.e. to add a constant correction, ${\rm Average}(\set{M})$-${\rm Average}(\set{B})$, to every number in~$\set{B}$. Let us call this mean-corrected set $\set{B}^*$. Similarly, we can correct both the mean and the variance -- we would just need two constants: one for the multiplicative and the other for the additive correction. This standard exercise produces a new set $\set{B}^{**}$ which now has the same mean and variance as $\set{M}$. We can go on matching moments in this way, but let us stop here and let us build the distributions $\{b_i^*\}$ and $\{b_i^{**}\}$ from $\set{B}^*$ and $\set{B}^{**}$ in the exact same way as we built the original $\{b_i\}$ from $\set{B}$.}
\begin{center}
    \rule[1ex]{.5\textwidth}{.5pt}
\end{center}
Let us now have a closer look at the Local Vol market. Replacing $\{b_i\}$ with $\{b_i^*\}$ and then with $\{b_i^{**}\}$ we see a drop in model risk premium from 6.06\% to 2.56\% and then to 1.53\% (per annum) -- see Fig.~1A. This means that over a half of model risk comes from the failure of the booking proposal to get the correct forward. Half of the remaining model risk comes from the erroneous assumption that the target variance is achieved perfectly. To understand the remaining 1.53\% we use Eq.~(\ref{Eq:f=b/m}) and plot $f_i^{**}=b_i^{**}/m_i$ (see the green line on Fig.~1B). We immediately recognize the risk-reversal shape of a skew product~\cite{Soklakov_2013a}. The vol-targeting scheme may reduce the skew but, clearly, does not eliminate it. We identify the failure of the booking proposal to capture skew as a major contribution to the residual MRP of 1.53\%.

It is instructive to ponder a little bit on the insight we just gained regarding the specific example of options on vol-targeted indices. Before doing the analysis, how many of us would confidently predict that the problems with the forward would dominate all other limitations combined? It might appear obvious to us now but, without the benefit of hindsight, how many practitioners would even mention forward as a potential problem? How many practitioners, when asked about top three limitations, would mention skew as the least important? Even in this very simple example we gained significant insight.

As a further illustration of Eq.~(\ref{Eq:f=b/m}) as an explanatory tool for the origins of risk, it is helpful to contrast Fig.~1B, which we just discussed, with the exact same analysis performed for Black-Scholes markets (see Fig.~1C). Such markets have no skew and, consequently, Fig.~1C provides a very different picture. We see a near-constant bond-like payoff for a wide range of values around ATM with some peaks in the far wings. This is a classic signature of Monte-Carlo noise. It is explained by good convergence across most of the grid of equally spaced buckets with poor sampling in the furthest buckets.

\section{Summary and Discussion}

Investments combine and contrast two kinds of knowledge: market-implied and investor-believed. Without any loss of generality, we can view both kinds of knowledge as models. Model risk assessment and investment design merge into a single field.

Using the specific example of options on vol-targeted indices, we showed how the methods of Quantitative Structuring allow us to measure and dissect model risk.
It is important to emphasize that the techniques we used are completely independent of the example. Equations~(\ref{Eq:f=b/m}) and~(\ref{Eq:MRR}) together with the generic moment-matching techniques are all we used to understand the origins and materiality of the risks.

What would be different if, instead of model choice, we decided to investigate, say, the quality of calibrations? Take, for instance, the SLV model and its ability to fit the vanilla markets. We would use terminal stock value as the variable, imply $\{m_i\}$ and $\{b_i\}$ from the market and the calibrated SLV respectively, and perform the analysis in the exact same manner as above. Similarly, we could have given an illustration of implementation risk -- using two different implementations of the same model to compute $\{m_i\}$ and $\{b_i\}$. Different flavors of model risk, different context (such as financial products) -- all of that would be reflected in the constructions of the distributions $\{m_i\}$ and $\{b_i\}$.

In the following we further clarify the methodology by summarizing it as an easily-productionizable algorithm
\begin{enumerate}
    \item {\bf List all relevant variables}\\
    This is the opportunity to specify the type of products which are covered by the investigation. For example, vanilla equity options depend on the terminal equity values, $S_T$. Asians and lookbacks depend on $\sum_{t<T} S_t$ and $\max_{t<T} S_t$ respectively. Most auto-callables are derivatives of the pair $(S_\tau,\tau)$ where $\tau$ is the autocall time.
    \item {\bf Define buckets on the variables}\\
    This is the opportunity to capture the granularity of the relevant markets. Here we note the number of liquid strikes, barrier levels, etc. which might be traded at any time. For example, the above analysis of options on vol-targeted index used $N=20$ equally spaced buckets covering the entire Monte-Carlo range of simulated $X_{\rm 6m}$. If instead we looked at vanillas on liquid equity indices, typical values for $N$ would be in hundreds.
    \item {\bf Imply probability distributions}\\
    This can be done either directly from the Monte-Carlo engine as illustrated above or indirectly -- via pricing of digitals or vanillas. The latter method is important when using PDE-based pricers especially when working with early-exercise derivatives~\cite{Soklakov_2013a}.
    \item {\bf Make a judgement on materiality}\\
    Compute ${\rm MRP}$ and make a judgement whether, in your business area, this is a material rate of return.
    \item {\bf Locate Model Risk}\\
    Dissect and analyze model risk using the combination of the moment matching technique and the visual display of the payoff (as demonstrated in this paper).
\end{enumerate}
As a final remark, it is worth noting that despite the clear financial context of this paper, the above techniques can be considered alongside general techniques of statistics. Indeed, we could have just as easily chosen to look at risk factors which, in terms of their modeling, are not financial: life expectancies, rates of traffic accidents, weather readings, etc. It would be very exciting to see financial intuition, such as recognizing the shape of payoff functions, being used outside of finance. This could be a chance for our industry to contribute back to the wider scientific community.

\section{Appendix}
\subsection{Stochastic local volatility model}\label{sec:SLV}
The stochastic local volatility model used in this paper, ${\rm SLV}^{\,\,\rho}_{\eta/\kappa}$, is very similar to the one discussed in~\cite{Madan_2007}. The model is defined by two correlated diffusions
        \begin{equation}
            dS=\mu S\,dt+\sigma(S,t)Z(t)S\,dW_S\,,
        \end{equation}
        \begin{equation}
            d\ln Z=(-\kappa\ln Z-\eta^2/2)\,dt+\eta\,dW_Z\,,
        \end{equation}
        \begin{equation}
            dW_S\,dW_Z=\rho\,dt\,.
        \end{equation}
        Without the stochastic multiplier $Z$ this would be just the standard local volatility model. Additional parameters include vol-of-vol $\eta$, mean-reversion $\kappa$, and the spot-vol correlation $\rho$. The local volatility function, $\sigma(S,t)$, is built as described in~\cite{Madan_2007} to match vanilla markets.

\subsection{Investment performance}\label{sec:Performance}

Each and every investment idea is naturally followed by the question of its performance. In Quantitative Structuring every investment is built as a solution to an optimization problem which captures the goals of the client. In such a setting every investment comes with its own natural performance measure -- the objective of the optimization.

Imagine, for example, a market of options on some underlying $x$. Without any practical loss of generality, we can assume that $x$ takes discrete values and that the market consists of binary securities each paying 1 for some value of $x$ and zero otherwise. In real applications (such as SPX options markets) different values of $x$ stand for price intervals for the underlying (such as S\&P 500 index) and the number of different values of $x$ is determined by how many such intervals are resolved by the market, i.e. by the number of liquidly traded strikes.

A growth-optimizing investor into the above market is defined as someone who seeks a product with a payoff function $g_x$ which maximizes the expected rate of return
\begin{equation}\label{Eq:GrowthOptimization}
{\rm ER}=\sum_xb_x\ln g_x\ \ \ {\rm subject\ to\ the\ budget\ constraint}\ \ \ \sum_x g_x p_x=1.
\end{equation}
The meaning of quantities $p_x$ and $b_x$ is given by the above optimization: $b_x$ is the investor-believed probability that the market variable will fix at the value $x$, and $p_x$ is a market price for the security which pays 1 if that happens and zero otherwise. From the point of view of a growth-optimizing investor, ER is clearly the best and the most natural measure of investment performance.

In the case of more general investors, this is of course not true. It turns out, however, that the solution to~(\ref{Eq:GrowthOptimization}) is a very convenient tool for understanding general investors~\cite{Soklakov_2014_WQS}. Despite the very well known limitations, growth rate remains an intuitive and frequently used measure of performance. Indeed, it is probably the only number which is guaranteed to feature in performance reports on all scales: from marketing brochures of tiny start-ups to expert projections of the World Economy. This is due to a very simple practical fact -- most people who ever had a business, or even just a bank account, can tell whether $r$\% per annum is a good or a bad return for them. Furthermore, people have strong feelings on the strategy of growth optimization~\cite{Soklakov_2014_WQS}, and this adds further depth to their individual judgements including informed demand for further research.

In summary, performance of every investment should ideally be measured relative to its goals. Although theoretically pleasing, this is often difficult in practice. Quantitative Structuring suggests a growth-optimizing strategy as a useful practical benchmark.

Solving the optimization~(\ref{Eq:GrowthOptimization}), we compute
\begin{equation}
g_x=b_x/p_x\,.
\end{equation}
Intuitively, the prices $\{p_x\}$ must have something to do with the probabilities of $x$. They are, however, actual market prices and would not necessarily sum up to one. There are two competing reasons for that. First, there is time value of money which normally has the effect of reducing (discounting) the numerical values of prices. Second, every business has to make a living, so market makers charge commission on top of theoretical values. In summary, we write
\begin{equation}\label{Eq:factorization}
\sum_xp_x={\rm DF}\cdot {\rm CF}\,,
\end{equation}
where we introduced the discount and the commission factors DF and CF to capture the two competing effects. We introduce the market-implied probability distribution $m_x$ by normalizing the prices
\begin{equation}
m_x\stackrel{\rm def}{=}\frac{p_x}{{\rm DF}\cdot {\rm CF}}\,.
\end{equation}

Note that, given the prices $\{p_x\}$, the market-implied distribution $m_x$ is always well-defined. The particular numerical split of the normalization factor into DF and CF reflects the particular funding arrangements secured by the market makers. We define
\begin{equation}
{\rm RFR}\stackrel{\rm def}{=}-\ln {\rm DF}\,,\ \ \ \ {\rm CR}\stackrel{\rm def}{=}\ln {\rm CF}\,,
\end{equation}
where, to give some names, we chose RFR and CR as abbreviations of \lq\lq risk-free rate'' and \lq\lq commission rate'' respectively.

Putting everything together we compute the structure of the growth-optimal investment
\begin{equation}
g_x\propto f_x\stackrel{\rm def}{=}b_x/m_x
\end{equation}
and its performance
\begin{equation}
{\rm ER}=\sum_{x}b_x\ln f_x+{\rm RFR}-{\rm CR}\,.
\end{equation}
The value of ER can be computed for any investment. By construction, the growth-optimal investment has the highest possible value of ER. Additional risk-aversion~\cite{Soklakov_2013b} will of course reduce ER and the amount of reduction would quantify the price of risk-aversion in terms of expected return. Similarly, we can quantify the price of any other modification to the investment -- from finishing touches and ad-hoc adjustments to changes in investments' most fundamental assumptions. Model risk analysis considered in this paper is a modest example of the latter. For a completely different example, the interested readers are referred to~\cite{Soklakov_2014EqPuzzle}.



\begin{thebibliography}{1}
\bibitem{Soklakov_2011} Soklakov, A., \lq\lq Bayesian lessons for payout structuring", RISK, Sept. (2011), 115-119. arXiv:1106.2882.

\bibitem{Soklakov_2014_WQS} Soklakov, A., \lq\lq Why quantitative structuring?", July (2015). arXiv:1507.07219.

\bibitem{Dupire_1994} Dupire, B., \lq\lq Pricing with a smile'', Risk, January (1994), 18-20.

\bibitem{DermanKani_1994} Derman, E. and Kani, I, \lq\lq Riding on a smile'', Risk, February (1994), 32-39.

\bibitem{Madan_2007} Ren, Y., Madan, D., Qian, M. Q., \lq\lq Calibrating and pricing with embedded local volatility models", RISK, September (2007).

\bibitem{Kelly_1956} Kelly J L Jr, \lq\lq A New Interpretation of Information Rate", Bell System Technical Journal, 917-26 (1956).

\bibitem{Soklakov_2013a} Soklakov, A., \lq\lq Deriving Derivatives", April (2013). arXiv:1304.7533.

\bibitem{Soklakov_2013b} Soklakov, A., \lq\lq Elasticity theory of structuring", April (2013). arXiv:1304.7535.

\bibitem{Soklakov_2014EqPuzzle} Soklakov, A., \lq\lq Quantitative Structuring vs the Equity Premium Puzzle", July (2015). arXiv:1507.07214.

\end{thebibliography}
\end{document}